\documentclass[twocolumn,10pt]{article}

\pdfoutput=1

\usepackage{graphicx}
\usepackage{amsmath}
\usepackage{amssymb}
\usepackage{fullpage}
\usepackage{theapa}

\newcommand{\bea}{\begin{eqnarray}}
\newcommand{\eea}{\end{eqnarray}}
\newcommand{\ben}{\begin{enumerate}}
\newcommand{\een}{\end{enumerate}}

\begin{document}

\sloppy

\title{RETAIN: a Neonatal Resuscitation Trainer \\ Built in an Undergraduate Video-Game Class}

\author{%
Vadim Bulitko \\
\small Dept. of Computing Science, University of Alberta\\
\small Edmonton, Alberta, T6G 2E8, CANADA \\
\small {\tt bulitko@ualberta.ca}
\and
Jessica Hong \\
\small Dept. of Art and Design, University of Alberta\\
\small Edmonton, Alberta, T6G 2E8, CANADA \\
\small {\tt jhong5@ualberta.ca}
\and
Kumar Kumaran \\
\small Royal Alexandra Hospital Site NICU \\
\small 10240 Kingsway Avenue NW \\
\small Edmonton, Alberta, T5H 3V9, CANADA \\
\small {\tt Kumar.Kumaran@albertahealthservices.ca}
\and
Ivan Swedberg \\
\small Dept. of Psychology, University of Alberta\\
\small Edmonton, Alberta, T6G 2E8, CANADA \\
\small {\tt iswedber@ualberta.ca}
\and
William Thoang \\
\small Dept. of Computing Science, University of Alberta\\
\small Edmonton, Alberta, T6G 2E8, CANADA \\
\small {\tt thoang@ualberta.ca}
\and
Patrick von Hauff \\
\small Faculty of Medicine \& Dentistry, University of Alberta\\
\small Edmonton, Alberta, T6G 2E8, CANADA \\
\small {\tt vonhauff@ualberta.ca}
\and
Georg Schm\"{o}lzer \\
\small Centre for the Studies of Asphyxia and Resuscitation \\
\small Neonatal Research Unit, Royal Alexandra Hospital \\
\small 10240 Kingsway Avenue NW \\
\small Edmonton, Alberta, T5H 3V9, CANADA \\
\small {\tt Georg.Schmolzer@albertahealthservices.ca}
}

\date{July 2, 2015}

\maketitle

\begin{abstract}
	Approximately ten percent of  newborns require some help with their breathing at birth. About one percent require extensive assistance at birth which needs to be administered by trained personnel. Neonatal resuscitation is taught through a simulation based training program in North America. Such a training methodology is cost and resource intensive which reduces its availability thereby adversely impacting skill acquisition and retention. We implement and present RETAIN (REsuscitation TrAIning for Neonatal residents) -- a video game to complement the existing neonatal training. Being a video game, RETAIN runs on ubiquitous off-the-shelf hardware and can be easily accessed by trainees almost anywhere at their convenience. Thus we expect RETAIN to help trainees retain and retrain their resuscitation skills. We also report on how RETAIN was developed by an interdisciplinary team of six undergraduate students as a three-month term project for a second year university course.
\end{abstract}

\section{Introduction}

Establishing breathing and oxygenation after birth is vital for survival and long-term health. The vast majority of newly born infants make the transition from intrauterine to extrauterine life without help. However, approximately $10\%$ of newborns require some form of assistance to breathe at birth and approximately $1\%$ require a more intensive resuscitation. Despite such care, approximately $900$ thousand newborn infants die annually worldwide due to birth asphyxia~\cite{spector2008}.

The delivery room is often a stressful environment where decisions are made quickly and resuscitators need to have good cognitive, psychomotor and communication skills. They also must have good team-management skills. However, the ``coming together'' of all theses skills is often more difficult than is widely appreciated. The neonatal training paradox~\cite{peter2005} describes the infrequent occurrence of neonatal emergencies with the risk of clinicians being ``unprepared, hesitant, and highly anxious''~\cite{aron2009} due to the lack of practical learning experiences. Such high acuity, low occurrence (HALO) situations arise infrequently but still require a high level of cognitive and technical competency. Neonatal and infant resuscitation perfectly meet the description of such HALO events~\cite{lou2008}. Such events lend themselves well to simulation-based training~\cite{aron2009}. Simulation-based medical education (SBME) has been identified as ``a highly effective instructional strategy for the acquisition and retention of skills requisite to competent performance in dynamic, high-pressure, high-consequence environments'' such as delivery rooms~\cite{jodee2011}.
The \citeauthor{jc2004} \citeyear{jc2004}, reporting on preventing infant death and injury during delivery, highlighted that inadequacies in proficient neonatal resuscitation account for over two thirds of mortality and morbidity.

\begin{figure}[t]
	\begin{center}
	\includegraphics[width=\columnwidth]{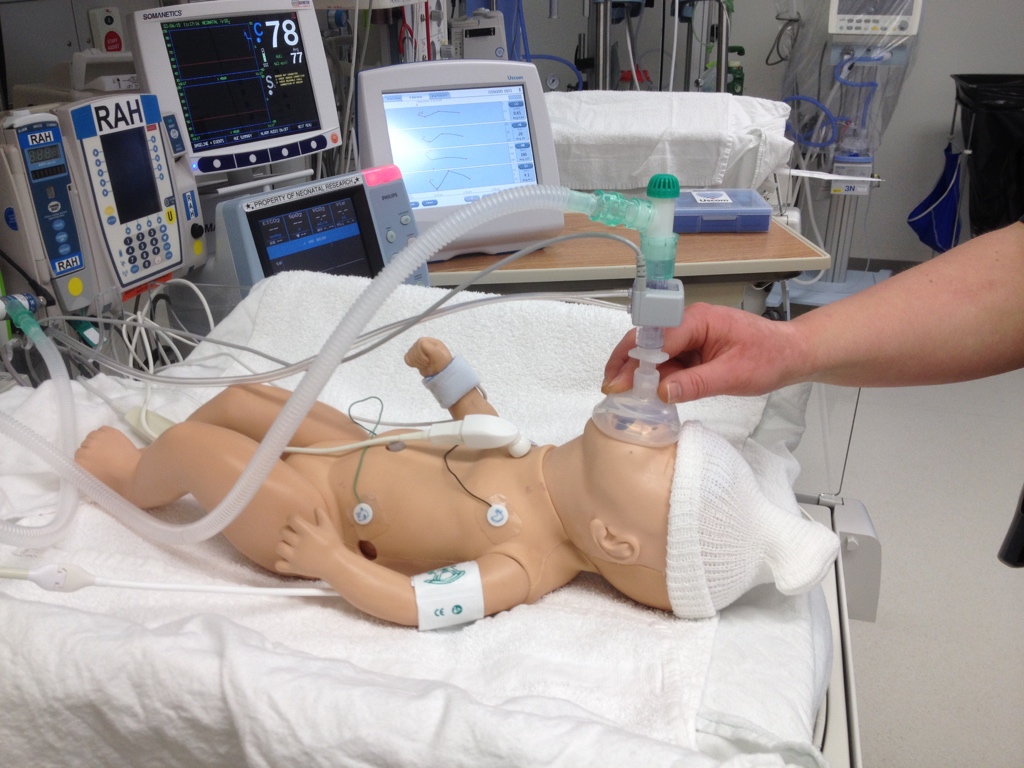}
	\end{center}
	\vspace{-0.4cm}
	\caption{Simulation-based training on neonatal resuscitation using a mannequin.}
	\label{fig:smbe}
\end{figure}

An international consensus on resuscitation emphasizes SBME to enhance performance in real-life clinical situations~\cite{pmid20956231}. However, the current neonatal training using a mannequin (Figure~\ref{fig:smbe}) requires access to a specialized facility, which needs to be equipped and staffed with instructors trained in SBME and debriefing. Consequently, SBME in its current form is time and cost intensive and therefore not offered routinely or in all healthcare facilities~\cite{pmid25153910}. Accordingly, the current certification requirements demand only a single four-hour SBME refresher course every two years to remain certified in neonatal resuscitation. While SBME has been shown to  improve performance initially after training~\cite{pmid9827341,pmid15276892},  both cognitive and technical skills significantly deteriorate within months~\cite{pmid25153910}. Although the optimal frequency for a refresher training is unclear, evidence suggests performing at least biannual training to ensure maintenance of knowledge and skills~\cite{pmid25153910}.

Using a video-game-like computer simulation to train various aspects of neonatal resuscitation has several advantages, all of which are critical when training healthcare professionals: controlled and risk-free learning, repetitive deliberate practice, customizability of training experiences to individual needs, objective assessment of and feedback on trainees' performance, easy accessibility and a less stressful environment. Although SBME can provide all of the above short of easy accessibility, there are potential disadvantages including a biased or distracted instructor or a single instructor who might be unable to observe everything during a team scenario. Using a computer-based simulation the instructor can receive a computer printout (or log) at the end of the game/simulation, which would allow him/her to identify what went well and what did not and debrief the trainee accordingly. Thus, in this paper, we adopt the approach of using a computer-based simulation game~\cite{sitzmann2011meta} to compliment physical simulation-based training on neonatal resuscitation. 

The primary contribution of this paper is a presentation of such a video game, called {\em REsuscitation TrAIning for Neonatal residents} (RETAIN). We describe the structure of the game and the non-playable characters driven by Artificial Intelligence (AI) that the trainee interacts with. RETAIN was developed by a team of six undergraduates as a term project for a second-year university class. Thus the second contribution of this paper are lessons learned from the production. With the proliferation of university-level courses and programs in video game development, we believe this account will be useful to other institutions.

The rest of the paper is organized as follows. We define the problem of neonatal resuscitation training with a serious game~\cite{Michael:2005:SGG:1051239} in Section~\ref{sec:problemFormulation}. Section~\ref{sec:relatedWork} reviews existing work relevant to the problem. We introduce the design of our approach (RETAIN) in Section~\ref{sec:design} and describe the production process in a video-game class in Section~\ref{sec:production}. In Section~\ref{sec:empiricalEvaluation} we present a preliminary evaluation of the approach. We then conclude the paper with a discussion of future work in Section~\ref{sec:futureWork}.

\section{Problem Formulation}
\label{sec:problemFormulation}

While, generally speaking, neonatal resuscitation training encompasses technical motor skills (e.g., bag and mask ventilation), decision-making skills (e.g., following the correct resuscitation procedure), communication and teamwork, we focus on decision-making skills. Studies have shown that not following a correct resuscitation procedure is responsible for approximately $60$-$70\%$ of all failures in the task~\cite{pmid23866717,pmid25125582}.  

The current physical simulation-based training is initially effective~\cite{pmid22594362} but frequent refresher training sessions are necessary for the trainee to retain the decision-making skills~\cite{pmid9827341,pmid15276892}. With the current physical training methodologies frequent refresher sessions are cost-prohibitive. Thus, the problem we address in this paper is to create a cost-effective and easily accessible non-physical training system for neonatal resuscitation that can complement the current training regimes and enables more frequent refresher training sessions for clinical personnel.

\section{Related Work}
\label{sec:relatedWork}

Simulation, which originated from aviation and spaceflight training programs, was adopted by anesthesiologists in the 1960s which eventually led to the development of simulation-based medical education~\cite{abrahamson2004effectiveness,schwid1992anesthesiologists}. 
Medical education and, in particular, surgical training has been an active area for the development of both simulations and serious games~\cite{graafland2012systematic,kapralos2014overview}. 
In the following sections, we review games relevant to the current work.

{\em e-Baby}~\cite{Fonseca2015-ew} is a serious game in which players perform clinical assessment of oxygenation on preterm infants in a virtual isolette. The infants present a range of respiratory impairments from mild to serious. The players were provided with patient history and had to select appropriate tools for clinical assessment. The assessment was made by responding to a series of questions in a multiple-choice format. The questions drove the interaction and served as an assessment of the trainee's knowledge. The game was evaluated by nursing students who had free access to the simulation and was rated highly for its ease of use and as overall efficacy of learning. The goal of {\em e-Baby} was  acquisition of procedural knowledge pertaining to the clinical assessment. Our goal is to create a game that trains medical personnel on the application of pre-existing knowledge of clinical intervention (resuscitation) in stressful conditions.

LISSA~\cite{wattanasoontorn2013serious,wattanasoontorn2013kinect,wattanasoontorn2014lissa} is a serious game to teach cardiopulmonary resuscitation (CPR) and use of an automated external defibrillator. Players must perform CPR procedures in the correct order within a specified time limit. The system supports play and authoring modes. Emergency scenarios are authored from a predefined set of elements, and can be complemented with expositional material (e.g., a demonstration of how to apply CPR). Scenarios are modeled as finite state machines corresponding to a CPR flowchart. LISSA was evaluated with $60$ learners with no background in CPR, and four CPR instructors. Although it was found to lead to lower learning outcomes compared to conventional instruction alone, LISSA was shown to have a higher efficacy when used to complement mannequin-based instruction.
Although relevant to our problem, LISSA differs in a number of key aspects: it targets adult cardiopulmonary resuscitation rather than neonatal resuscitation, and is intended for a general audience rather than clinical trainees. LISSA also aims to teach motor skills via a use of Kinect~\cite{wattanasoontorn2013kinect} which is beyond the scope of our problem (decision-making skills).

\citeauthor{kalz2013design} \citeyear{kalz2013design} report on the development of a mobile game-based resuscitation training for first responders. The goal of the project was to augment, rather than to replace, face-to-face training. Specifically, the authors sought to ``increase procedural knowledge, train processes in an emergency situation and to influence willingness to help and self-efficacy''~\cite{kalz2013design}. As with LISSA the intended audience were laypeople and the domain was adult CPR. 

RELIVE~\cite{loconsole2015relive} and {\em Viva!}~\cite{semeraro2014relive} were developed for use by general audiences as part of an awareness week around CPR training~\cite{ristagno2014achievements}. RELIVE was intended as a low-cost trainer for non-clinical use and employed Microsoft Kinect, a realistic 3D environment, and game-like interface to provide feedback to players about their CPR performance. 
{\em Viva!} is a serious game intended to raise awareness among adults and children of CPR training. Accordingly, the game employed a 2D retro-cartoon style. A variety of rescue scenarios take place in different simulated locations. Players perform CPR by clicking on icons representing actions of interest, in order to save characters from cardiac arrest. The game can be played in two modes. In story mode, players are led through structured training and must achieve a high level of performance before they can proceed. In tournament mode, players are able to engage in ready-to-play emergency scenarios, to test the accuracy of their CPR performance. Players may also challenge friends to compete on the accuracy of their maneuvers. As with previous efforts and unlike for the problem we are solving, {\em Viva!}'s target audience was laypeople.

{\em Pulse!!} is a virtual clinical-training environment for trauma management~\cite{johnston2005pulse,mcdonald2011pulse,breakaway2008}. Aimed at clinical professionals and learners, {\em Pulse!!} allows players to train clinical skills in numerous emergency situations. Players interact with patients in a highly realistic clinical environment, furnished with a variety of medical equipment and player-controllable staff members. The system includes a case-authoring tool, scene editor, tutoring system, and asset library of characters, environments, equipment and physiological processes. In an assessment of its effectiveness as a learning and assessment tool, {\em Pulse!!} was found to be significantly more effective than paper-based learning, and was rated highly engaging by players~\cite{mcdonald2011pulse}. While {\em Pulse!!} was designed to be a comprehensive simulation environment for adult trauma, the problem we are addressing is resuscitation in newborns. 

{\em Triage Trainer}~\cite{knight2010serious} is a serious game designed to teach major incident triage to clinical professionals. Developed to be played on a desktop or a laptop computer, the game allows its players to practice triage (prioritizing which patients to treat when) in a realistic immersive 3D environment. Players navigate and interact with casualties using the mouse and keyboard. Assessment is done by clicking on a series of icons representing various examinations (e.g., breathing check, pulse rate check) and manipulations (e.g., open airway, tag a casualty with triage rating). The focus of the game is on rapid execution of process-based knowledge. The authors found that participants who played the game had significantly greater accuracy on a triage task than did participants who took part in the control activity (card sort). Although it addresses clinical decision-making under pressure, {\em Triage Trainer} deals with the domain of mass casualty triage, not neonatal resuscitation.

{\em Surgical Improvement of Clinical Knowledge Ops} (SICKO) is a web-based game designed to practice and assess clinical decision-making in surgery~\cite{lin2015validity}. The game was inspired by and developed from the original work on {\em Septris}~\cite{wykes2012game}, a web-based game to teach learners about sepsis.
The purpose of SICKO is to simulate decision-making under pressure rather than psychomotor skills. In the game, players must balance the care of multiple patients, as they would in real life. As they do so, Dr. Sicko, represented by a cartoon figure, shows his approval by smiling or frowning and adding or deducting points from the player's score. Scenarios cover a range of acuity and complexity and can include X-ray and MRI imaging. Similar to other work mentioned in this section, SICKO addresses clinical decision-making under pressure, however, the clinical environment is highly abstracted. Inspired by {\em Tetris}~\cite{tetris}, the patients are represented as faces which descend from the top of the screen in columns, and must be successfully treated before reaching the bottom. Additionally, the domain addressed by SICKO does not encompass neonatal resuscitation.

\section{RETAIN Design}
\label{sec:design}

\begin{figure}[t]
	\begin{center}
	\includegraphics[width=\columnwidth]{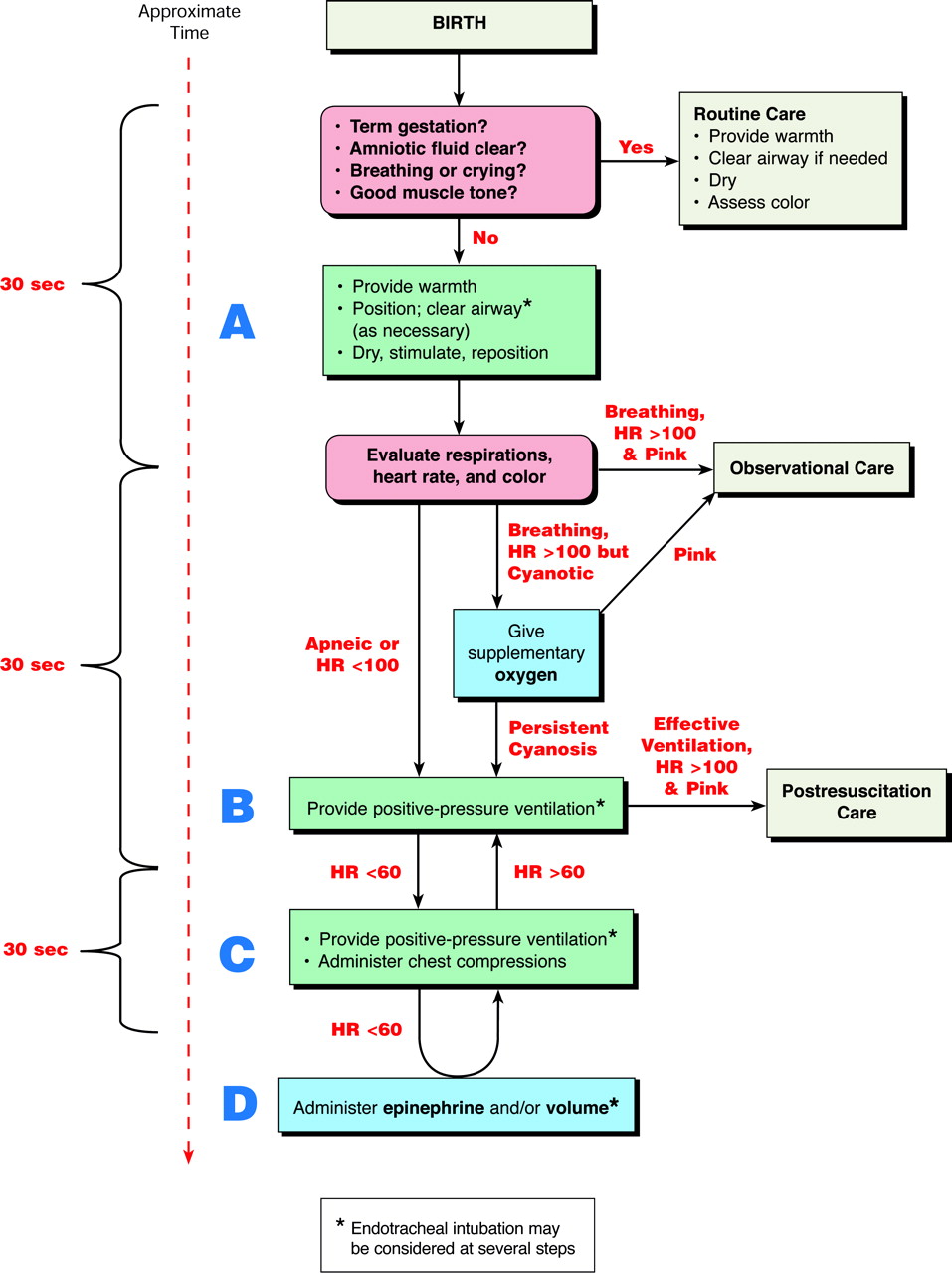}
	\end{center}
	\vspace{-0.4cm}
	\caption{The neonatal resuscitation algorithm by the International Liaison Committee on Resuscitation.}
	\label{fig:chart}
\end{figure}

\begin{figure}[t]
	\includegraphics[width=\columnwidth]{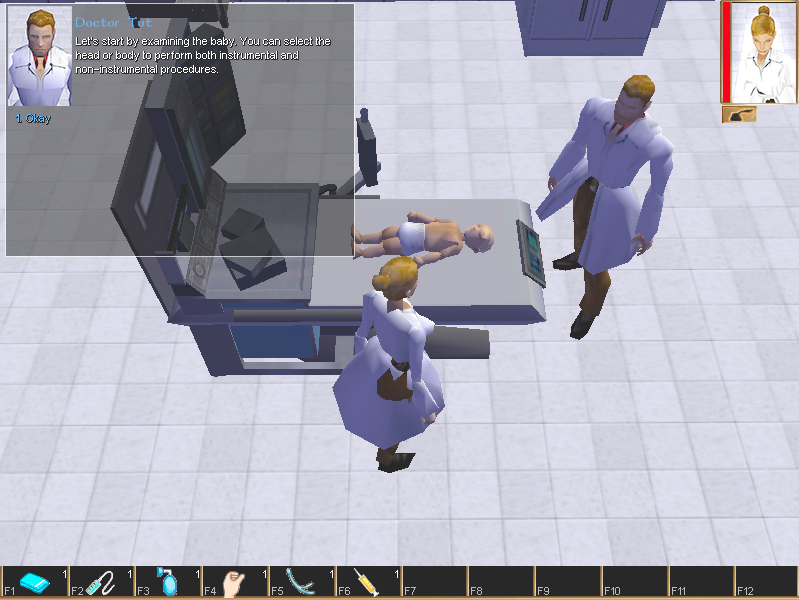}
		\vspace{-0.4cm}
	\caption{The game interface.}
	\label{fig:interface}
\end{figure}

We designed RETAIN around the resuscitation flow chart shown in Figure~\ref{fig:chart}, reproduced from~\cite{pmid20956231}. To ramp the decision-making complexity gradually, we built the game with four levels of increasing difficulty. An introductory (zero-th) level served as a tutorial of the game interface where the trainee, controlling a medical resident, followed directions of a doctor experienced in resuscitation. The doctor was a non-playable character controlled by AI whose dialogue structure was implemented as a dialogue tree~\cite{aurora}. Recorded voice overs were used in conjunction with written dialogue to have the doctor converse with the trainee (Figure~\ref{fig:interface}). The trainee interacted with the game with a keyboard and mouse for level navigation and menu selections.

\begin{figure}[t]
	\includegraphics[width=\columnwidth]{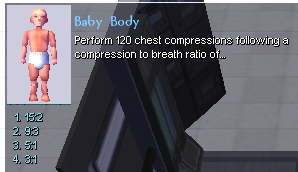}
	\vspace{-0.4cm}
	\caption{Chest compression parameter choices.}
	\label{fig:chestCompression}
\end{figure}

The doctor in the tutorial level gave a complete guidance on which actions to take and served as an introduction to the interface. Having passed the tutorial, the trainee would tackle three resuscitation scenarios of different complexity. While in each of them the trainee was still paired with an AI-controlled doctor, he or she now received incomplete guidance from the AI-controlled doctor and had to use their own judgment to select the right action at the right time. Furthermore, most actions now had a parameter (e.g., the frequency of chest compressions) which had to be specified by the trainee as well (Figure~\ref{fig:chestCompression}). 

\begin{figure}[t]
	\begin{center}
	\includegraphics[height=2.2cm]{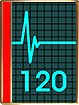}\hspace{0.2cm}
	\includegraphics[height=2.2cm]{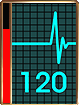}\hspace{0.2cm}
	\includegraphics[height=2.2cm]{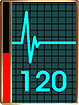}\hspace{0.2cm}
	\includegraphics[height=2.2cm]{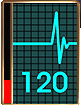}
	\end{center}
		\vspace{-0.4cm}
	\caption{The baby's health represented with a health bar next to the baby's pulse.}
	\label{fig:healthBar}
\end{figure}

An incorrect action selection (e.g., applying suction instead of performing chest compressions) or an incorrect parameter value (e.g., using the chest compressions with the breath ratio of $5$:$1$ instead of $3$:$1$) counted as a mistake and the trainee received an immediate feedback in the form of auditory cue (a bell tone) and an utterance from the doctor. Each mistake also decreased the baby's health level (shown as a vertical bar in Figure~\ref{fig:healthBar}). After four mistakes the baby died, ending the scenario. By committing fewer than four mistakes the trainee saved the baby, also ending the scenario.

\begin{figure}[htbp]
	\begin{center}
	\includegraphics[width=\columnwidth]{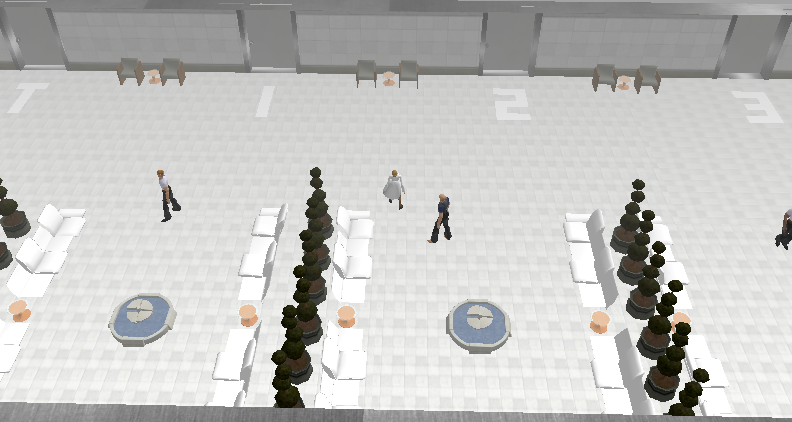}
	\end{center}
	\vspace{-0.4cm}
	\caption{The hospital lobby served as the game hub.}
	\label{fig:lobby}
\end{figure}

\begin{figure}[t]
	\begin{center}
	\includegraphics[width=\columnwidth]{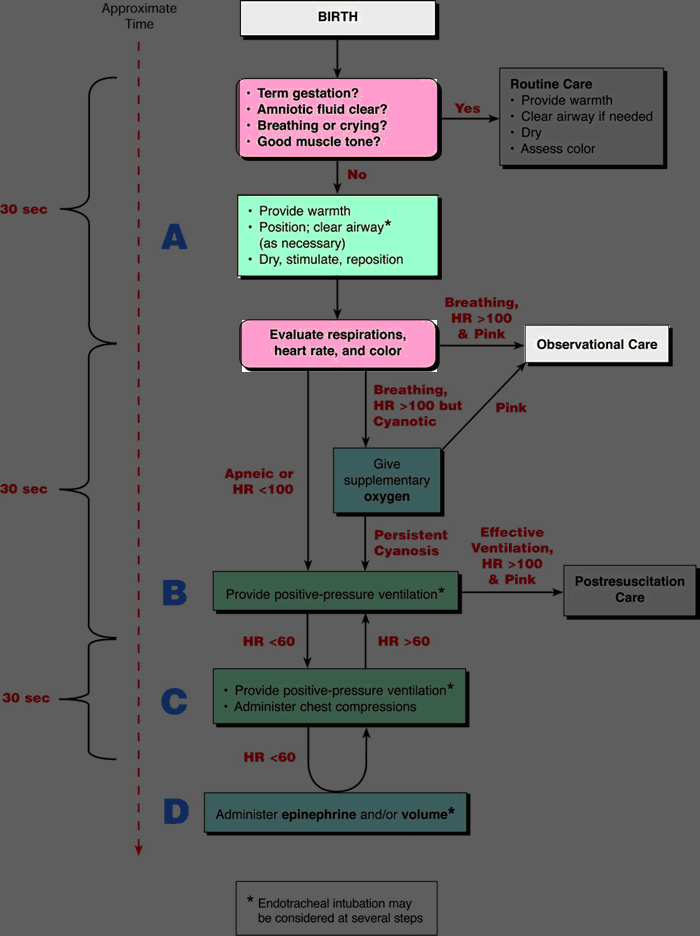}
	\end{center}
	\vspace{-0.4cm}
	\caption{Parts of the resuscitation algorithm covered by training scenario 1.}
	\label{fig:chart1}
\end{figure}

\begin{figure}[t]
	\begin{center}
	\includegraphics[width=\columnwidth]{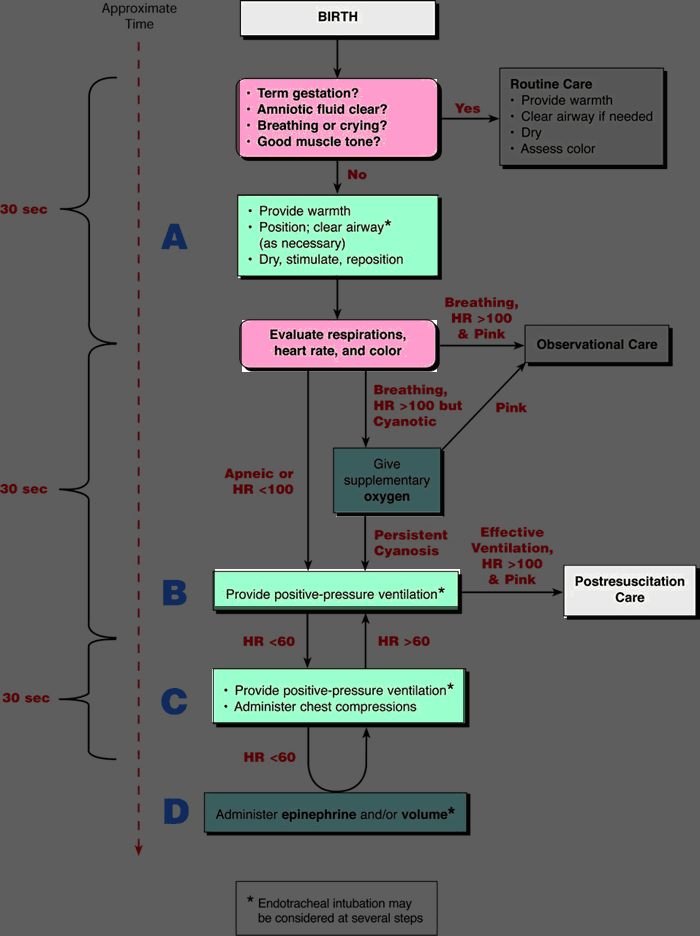}
	\end{center}
	\vspace{-0.4cm}
	\caption{Parts of the resuscitation algorithm covered by training scenario 2.}
	\label{fig:chart2}
\end{figure}

\begin{figure}[t]
	\begin{center}
	\includegraphics[width=\columnwidth]{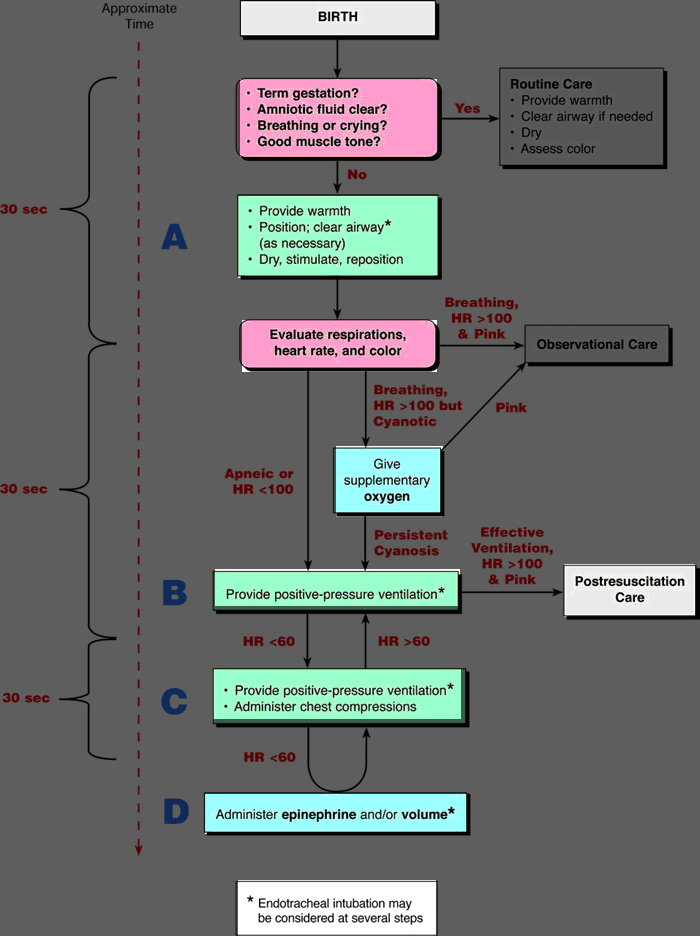}
	\end{center}
	\vspace{-0.4cm}
	\caption{Parts of the resuscitation algorithm covered by training scenario 3.}
	\label{fig:chart3}
\end{figure}

The trainee tackled each of the three scenarios by accessing one of the three doors from a hub area representing the hospital's lobby (Figure~\ref{fig:lobby}). 
The later scenarios were more difficult as in order to save the baby they required the trainee to take (i) a longer action sequence ($6$, $9$ and $13$ actions for the three scenarios respectively) and (ii) to use a wider variety of actions ($5$, $7$ and $9$ out of a total of $9$ actions RETAIN supported). Thus, the trainee was tested on a progressively larger portion of the resuscitation algorithm (Figures~\ref{fig:chart1}, \ref{fig:chart2}, \ref{fig:chart3}). Note that while RETAIN supported a total of $9$ actions, most of them took a parameter thereby greatly increasing the number of choices available to the trainee to pick from.

\section{RETAIN Production}
\label{sec:production}

The entire production was carried out as a term-project in a second-year undergraduate course. The production took three months and involved a team of six undergraduate students from different disciplines: computing science, psychology, art and design and biology. The team was advised by two neonatologists on a weekly basis. Each week the developer team also met with a former graduate of the course serving as a mentor. The team additionally received a 3D baby model from an external artist and had external voice actors record voice overs for in-game characters. 

The production faced a number of challenges. First, the development toolset used for the production, {\em Aurora Toolset}~\cite{aurora} is more suitable for dialogue-based isometric-view RPG adventures such as {\em Neverwinter Nights}~\cite{nwn}. The team spent a considerable amount of time modifying the interface for the medical simulation (e.g., a level map had to be displayed off screen). 
Second, the six students were from different departments/faculties and were enrolled in four to five other courses during the term. This made scheduling meetings difficult. The team ended up meeting both on- and off-campus, often on weekends. Third, none of the team members had previously worked on a term-long project with students with other backgrounds, nor had any team member previously developed a video game. Timely communications were critical and were organized through on-line collaboration tools including Google apps, Trello, Skype and Facebook.

\section{Preliminary Evaluation}
\label{sec:empiricalEvaluation}

To date, RETAIN has gone through several evaluation phases. Early vertical slices of the game were tested by the team’s mentor and the course instructor. A beta version was evaluated by the course teaching assistants (two computing science graduate students) and the classmates (thirty students). The final version was evaluated by the course instructor and by eight members of an award committee. The committee was tasked with selecting award winners for an annual game-award ceremony tied to the course and consisted of game developers (both AAA and independent) as well as academics. Not only the committee recognized RETAIN with an award but also the committee members, including experienced AAA commercial game developers, reported unexpected stress while playing the game due to the gravity of the context (saving the baby). This was in contrast to the other nominated games which were all made for entertainment purposes. Finally, the game was evaluated by two neonatologists who felt that overall the game displayed important aspects of basic neonatal resuscitation training. In particular, they deemed that the game was able to keep the player engaged and involved, with sufficient drama and stress, while providing a platform for learning by doing in a cost-effective and engaging fashion.

\section{Future Work}
\label{sec:futureWork}

While anecdotal, the evidence presented in the previous section is encouraging and opens up several directions for future work. The immediate next step is to evaluate RETAIN in a medical training environment. Preparations are underway to run user studies later this summer and in early fall. Following such an evaluation, a research-grade version of RETAIN will be developed. The future version will attempt to actively model the trainee's skills in neonatal resuscitation. This will allow an AI manager to tailor each training scenario to a given trainee. Specifically, matching the trainee's inferred skills against complexity of various scenario modules will allow the AI manager to estimate the trainee's degree of flow~\cite{flowStorytelling2015}. By keeping the trainee in a state of flow his or her learning may be improved~\cite{flow1990}. Being in a state of flow is also intrinsically rewarding and thus may encourage a trainee to practice more with the system. 

With a longer development cycle and a larger development budget, the next version of RETAIN will be built with a modern toolset, support a wider range of resuscitation scenarios and achieve a greater visual and medical fidelity. Finally, we are planning to implement an AI-driven feedback system which will critique trainee’s actions and score their overall performance.

\section{Conclusions}
\label{sec:conclusions}

Neonatal resuscitation is a crucial life-saving procedure in modern hospitals. While traditional training requires access to a costly specialized facility, we conjecture that complementing it with a video-game-based, easily accessible low-cost version will add training value via more frequent training sessions. In this paper we reported on a three-month undergraduate production of the first version of such a training game carried out as a term project in a second-year undergraduate class. The initial prototype and its development appear successful and encourage a development of a research-grade follow up.

\section{Acknowledgments}
We would like to thank other members of the RETAIN development team: Erik Estigoy, Connor Hastey Palindat, Vishruth Kajaria and Derek Kwan. Baby modeling was performed by Glenn Meyer. Voice actors were Fuad Sakkab, J.D. Macnutt, Nathan Wakeman, Gerard Capiuk and Jessica Hong. We also thank the class teaching assistants Yathirajan Brammadesam Manavalan and Sergio Poo Hernandez. We also appreciate the support from the Community Service Learning centre at the University of Alberta and the Heart and Stroke Foundation of Alberta.

\bibliographystyle{theapa}
\bibliography{aiide,patrick5,patrick5b}

\end{document}